\documentclass[10pt,conference]{IEEEtran}
\IEEEoverridecommandlockouts
\usepackage{cite}
\usepackage{amsmath,amssymb,amsfonts}
\usepackage{algorithmic}
\usepackage{graphicx}
\usepackage{textcomp}
\usepackage[dvipsnames]{xcolor}
\usepackage{longtable}
\usepackage{todonotes}
\usepackage{multirow}
\usepackage{multicol}
\usepackage[many]{tcolorbox}   
\usepackage{colortbl}
\usepackage{float} 
\usepackage{placeins}

\usepackage{enumitem}
\usepackage{soul}
\usepackage{balance}

\newcommand{\heng}[1]{\textcolor{black}{{#1}}}

\newcommand{\qql}[1]{\textcolor{black}{{#1}}}

\tcbset{
    sharp corners,
    colback = white,
}                           
\renewcommand\st[1]{}
\newtcolorbox{boxD}{
    colback = {white}, 
    colframe = {black}, 
    boxrule = 0pt, 
    toprule = 1.5pt, 
    bottomrule = 1.5pt, 
    left=3pt,right=3pt, top=2pt,bottom=2pt
}
\def\BibTeX{{\rm B\kern-.05em{\sc i\kern-.025em b}\kern-.08em
    T\kern-.1667em\lower.7ex\hbox{E}\kern-.125emX}}
\begin{document}

\title{Automated, Unsupervised, and \qql{Auto}\st{Non}-parameterized Inference of Data Patterns and Anomaly Detection}

\author{\IEEEauthorblockN{Qiaolin Qin, Heng Li*\thanks{*Corresponding author.}, Ettore Merlo, Maxime Lamothe}
\IEEEauthorblockA{Dept. of Computer and Software Engineering \\
Polytechnique Montreal, Montreal, Canada \\
\{qiaolin.qin, heng.li, ettore.merlo, maxime.lamothe\}@polymtl.ca
}
}

\maketitle

\begin{abstract}
\label{abstract}
With the advent of data-centric and machine learning (ML) systems, data quality is playing an increasingly critical role for ensuring the overall quality of software systems.
Data preparation, an essential step towards high data quality, is known to be a highly effort-intensive process. 
Although prior studies have dealt with one of the most impacting issues, data pattern violations, these studies usually require data-specific configurations (i.e., parameterized) or use carefully curated data as learning examples (i.e., supervised), relying on domain knowledge and deep understanding of the data, or demanding significant manual effort. 
In this paper, we introduce RIOLU: Regex Inferencer \st{nOn}\qql{autO}-parameterized Learning with Uncleaned data. RIOLU is fully automated, \st{non-}\qql{automatically }parameterized, and does not need labeled samples. 
RIOLU can generate precise patterns from datasets in various domains, with a high F1 score of 97.2\%, exceeding the state-of-the-art baseline.
In addition, according to our experiment on five datasets with anomalies, RIOLU can automatically estimate a data column's error rate, draw normal patterns, and predict anomalies from unlabeled data with higher performance (up to 800.4\% improvement in terms of F1) 
 than the state-of-the-art baseline, \qql{even outperforming ChatGPT in terms of both accuracy (12.3\% higher F1) and efficiency (10\% less inference time). }
 A variant of RIOLU, with user guidance, can further boost its precision, with up to \st{37.371\%}\qql{37.4\%} 
 improvement in terms of F1
. 
Our evaluation in an industrial setting further demonstrates the practical benefits of RIOLU.
\end{abstract}


\section{introduction}
\label{sec:introduction}

Data are an essential part of modern software~\cite{sommerville2015software}. 
In particular, data-centric~\cite{carey2008data} or AI-enabled software systems~\cite{ozkaya2020really} incorporate massive data to provide querying or intelligent services to users. 
Many studies have discussed the importance of data quality in software engineering~\cite{pachouly2022systematic, croft2022data, shi2022we, croft2023data, xu2023data, davoudian2020big}: poor data quality can harm software engineering efforts. 
\qql{In particular, prior studies have stressed the importance of data quality assurance for the overall software quality~\cite{hu2024autoconsis, tange2023dclink, lwakatare2021experiences}. For example, a recent study discusses the data inconsistency issue in mobile apps which can confuse users and cause app quality degradation~\cite{hu2024autoconsis}.}
According to an industrial investigation by Stonebraker and Rezig, data preparation could take over 80\% of the development time 
when executed manually~\cite{stonebraker2019machine}. 
To reduce the labor cost of data preparation
, companies and researchers aim to find automatic ways to understand the quality and detect anomalies in their data. 

Data pattern-related anomalies (or pattern anomalies) are a major type of \qql{flat structural} data anomaly
~\cite{abedjan2016detecting, mahdavi2019raha} 
and are widely studied by researchers~\cite{abedjan2016detecting, visengeriyeva2018metadata, mahdavi2019raha, song2021auto}, 
given their hard-to-detect and hard-to-debug nature~\cite{song2021auto}. Unlike \qql{log anomalies~\cite{le2022log}\cite{du2017deeplog} and other time-series anomalies~\cite{li2020pyodds}\cite{lai2021tods}\cite{xu2018unsupervised} which describe abnormal system events or status and typically require feature sets for anomaly detection, 
data pattern anomalies describe anomalies in the data structures themselves which may arise in features. They}\st{Pattern anomalies occur when certain data records} \qql{occur when a record} 
violate\qql{s} pre-defined or implicit 
patterns~\cite{abedjan2016detecting, mahdavi2019raha}. 
Prior work~\cite{padhi2018flashprofile} defines data patterns as \textit{regular expressions (regexes) that succinctly describe the syntactic variations in the strings}. For example, a data pattern for a dataset of dates could be $\backslash d\{4\}$-$\backslash d\{2\}$-$\backslash d\{2\}$. 
While pattern-based \textbf{data profiling} focuses on data description with patterns, pattern-based \textbf{anomaly detection} leverages the patterns to detect anomalies (i.e., data instances that violate the data patterns)~\cite{breck2019data, schelter2018automating, raman2001potter}. 
However, two critical \textit{challenges} face automated data profiling and pattern anomaly detection: (1) How can we automatically infer data patterns without prior data format knowledge, given its infinite possibilities? (2) How can we ensure the health of the inferred patterns (i.e., avoid anti-patterns that cover anomalies themselves)? 
This work addresses these challenges and provides an automated, unsupervised, and \st{non}\qql{automatically}-parameterized solution for pattern inference and anomaly detection.



Prior work has proposed different approaches for pattern-based data profiling~\cite{breck2019data, padhi2018flashprofile, ilyas2018extracting} (i.e., pattern inference) and anomaly detection~\cite{raman2001potter, yu2023human, song2021auto, schelter2018automating}. 
These approaches use either \textit{declarative} (i.e., manually designed) or \textit{automatically inferred} regular expressions to capture the patterns in the data~\cite{song2021auto}.
Although declarative tools such as Google's TFDV~\cite{breck2019data} and Amazon's Deequ~\cite{schelter2018automating} are straightforward to use, manually designing the regular expressions for each data column is still very effort- and time-consuming~\cite{song2021auto}. 
To reduce human effort, researchers designed tools 
that can automatically draw patterns from the data and identify anomalies. 
For example, XSystem~\cite{ilyas2018extracting} split the structure into three layers of representation (i.e., branch, token, symbol) to incrementally extract regular expressions from a provided table. 
FlashProfile~\cite{padhi2018flashprofile} clusters the records by their syntactic similarity and assigns patterns for each cluster; it then filters anomalies by dropping low-frequency (e.g., less than 1\%) patterns, after synthesizing column patterns. 
Auto-Validate~\cite{song2021auto} treats pattern quality as a global notion and ensures the generalizability of a pattern by learning it from one column and testing its generalization ability on other similar columns. 
However, these tools still suffer from two shortcomings: (1) parameter configuration that requires domain expertise: for example, XSystem requires a predefined number of branches for pattern generation, which necessitates prior knowledge of the number of valid patterns
; (2) insufficient precision: for example, the similarity-based approach used by FlashProfile cannot distinguish records that are seemingly close to each other, leading to inaccurate patterns. 

To overcome the difficulty of parameter 
configuration and improve the precision, we propose RIOLU (Regex Inferencer \st{nOn}\qql{autO}-parameterized 
Learning with Uncleaned data), a pattern inference and anomaly detection approach 
that can be adaptively parameterized. Inspired by Auto-Validate~\cite{song2021auto}, RIOLU regards high-quality patterns as patterns that can cover adequate healthy samples; inspired by Potter's wheel~\cite{raman2001potter}
, RIOLU selects the most expressive and concise patterns. Our work uses statistical heuristics and K-Means clustering \st{to remove the need for parameters.}\qql{to automatically infer parameters, thus avoiding manual parameter setting. }
Moreover, the approach is built using a rule-based progressive structure which enhances the pattern quality. 
To assess the usefulness of RIOLU, we address the following two research questions (RQs) in this study:
\noindent\underline{\textbf{RQ1: How well can RIOLU profile data in different do-}} \\\underline{\textbf{mains?}} 
Deriving precise patterns is a prerequisite for pattern anomaly detection. This RQ aims to assess RIOLU's ability in deriving patterns for data description from all records 
(\textit{a.k.a.}, data profiling). Our results show that RIOLU outperforms FlashProfile, a state-of-the-art open-sourced data profiler\qql{, and GPT-3.5 Turbo, on data in various domains}.

\noindent \underline{\textbf{RQ2: How precisely can RIOLU detect pattern anoma- }} 
\\\underline{\textbf{lies?}} 
Pattern anomalies are common in uncleaned data. This RQ assesses RIOLU's ability to learn patterns and detect anomalies in such data. We observe that RIOLU can effectively detect pattern anomalies in public datasets and an industry setting, outperforming state-of-the-art baselines.

Our work makes the following main contributions: 
\begin{itemize}[leftmargin=*]
    \item 
    We proposed a fully automated approach, RIOLU, for data pattern inference and anomaly detection without supervision or parameter configuration. 
    \item We demonstrated that a human-guided variant of RIOLU (Guided-RIOLU) can further improves the performance of the fully automated version (i.e., Auto-RIOLU\footnote{We use the name Auto-RIOLU to refer to the fully automated version, in order to distinguish it from the Guided-RIOLU version.}). 
    \item We evaluated Auto-RIOLU on real-world commercial data, with practitioners 
    confirming its ability to generate precise patterns for data quality verification and anomaly detection. 
\end{itemize}

Practitioners and researchers can leverage RIOLU to understand their data and improve data quality, reducing their time and effort in data preparation or data quality assurance.
Our data and code are published in a replication package~\cite{riolu}.

\section{Background}
RIOLU targets pattern-based data profiling and anomaly detection. 
This section first defines \textit{patterns} in our context, followed by the definitions of the two pattern-based tasks. 

\label{sec:backgrounds}

\subsection{Patterns}
\label{sec:pattern_language}
Following Song and He~\cite{song2021auto}, we define patterns as structured sequences containing basic components. In English-based patterns, basic elements include upper letters, lower letters, digits, and symbols~\cite{huang2018auto, song2021auto, he2021auto}. 
According to Ilyas \textit{et al.}~\cite{ilyas2018extracting}, symbols serve as a natural delimiter in splitting the patterns. 
For example, \textit{2024-Mar-01} is a string containing all these pattern elements and can be represented as ``digit\{4\}symbol\{1\}upper\{1\}lower\{2\}symbol\{1\}digit\{2\}''. Such patterns can be used to match or describe a data domain. 

\begin{figure*}[!t]
  \centering
  \includegraphics[width=0.85\textwidth]{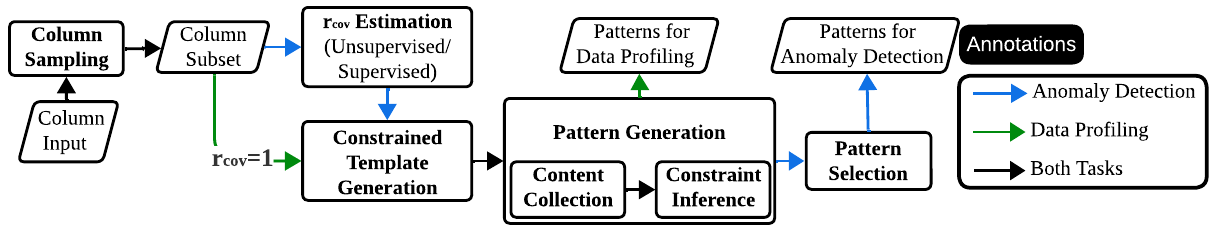}
  \vspace{-2ex}
  \caption{An overview of RIOLU's structure.} 
  \label{fig:methodology_overview}
  \vspace{-2ex}
\end{figure*}

\subsection{Pattern-Based Data Profiling}
\label{sec:pattern_based_data_profiling}
The goal of pattern-based data profiling is to distinguish patterns in the records by describing the data using derived regular expressions.
Compared to previous pattern-based data profiling tools such as Microsoft’s SQL Server Data Tools
(SSDT)~\cite{SSDTDocumentation} and Ataccama One~\cite{AtaccamaWebsite}, which heavily relies on a small predetermined set of atoms
, FlashProfile~\cite{padhi2018flashprofile} uses a bottom-top technique to cluster records through syntactic similarity and then extract patterns from each cluster; it requires no user-defined domains or base patterns (e.g., dates, phone numbers) to reach a high precision. However, we noticed that the tool may not be robust when the differences between records are subtle. For example, a column of state/province abbreviations should only contain two upper letters (i.e., [A-Z]\{2\}). Due to poor data quality, the column may also contain noisy records with lower letters (e.g., Qc, on) or incorrect string lengths (e.g., M, ABB). Given the small syntactic distances between different spelling types, they may be identified as one cluster, and the profiling result would be ``[a-zA-Z]+'', which fails to precisely describe the underlying pattern. 
\vspace{-0.2cm}
\subsection{Pattern Anomaly Detection}
\label{sec:pattern_anomaly_detection}

\qql{Pattern anomalies indicate records that violate pre-defined or inferred 
regex-like patterns~\cite{abedjan2016detecting}~\cite{mahdavi2019raha}. }Following Huang and He~\cite{huang2018auto}, we define pattern anomaly detection as the process of filtering inconsistent data (i.e., anomalies) using regular expressions. As introduced in Sec~\ref{sec:introduction}, existing tools provide automatic inference approaches for pattern anomaly detection. 
Although these tools have largely reduced human labor, they always have the problem of requiring domain-dependent parameter settings. For example, XSystem~\cite{ilyas2018extracting} requires the user's prior knowledge of the number of branches (i.e., number of acceptable pattern types), which could be hard to fix at the first use and may not be constant across domains (e.g.,  a column for ``id'' can only accept one branch as \textit{$\backslash$d\{5\}}, while a column for ``date'' may accept both patterns of \textit{$\backslash$d\{4\}-$\backslash$d\{2\}-$\backslash$d\{2\}} and 
 \textit{$\backslash$d\{4\}/$\backslash$d\{2\}/$\backslash$d\{2\}}, written as 2024-01-01 and 2024/01/01, respectively, as valid). 
FlashProfile~\cite{padhi2018flashprofile} 
require users to define the threshold for low-frequency patterns 
(i.e., a pattern with a frequency lower than the threshold is considered as abnomal). However, a threshold may not hold for different datasets, and it is difficult for users to determine a solid threshold without fully
comprehending the dataset. 

\section{Approach}
\label{sec:approach}

\subsection{Overview}
\label{sec:overview}

The input of RIOLU is a two-dimensional table with random numbers of columns and rows. The structure of each column is undefined: for example, they can be names, IDs, or URLs. The goal of RIOLU is to automatically derive the patterns (e.g., YYYY-MM-DD) of each column without prior knowledge; the patterns can then be used to detect data anomalies (e.g., invalid URLs). As shown in Fig~\ref{fig:methodology_overview}, the approach of RIOLU for determining patterns and detecting anomalies in a column can be decomposed into the following steps:
\begin{itemize}[leftmargin=*]
    \item \textbf{Column Sampling}: Sample a subset of data from the column to generate the patterns. 
    \item \textbf{Coverage Rate ($r_{cov}$) Estimation}: Estimate the percentage of healthy values ($r_{cov}$) in each column. 
    \item \textbf{Constrained Template Generation}: Generate raw templates for each record with 
    a granularity constraint. 
    \item \textbf{Pattern Generation}: Generate pattern constraints for each template according to the coverage rate.
    \item \textbf{Pattern Selection}: Select patterns based on some heuristics (e.g., their generalizability). 
\end{itemize}

As shown in Fig~\ref{fig:methodology_overview}, all five steps are needed for the anomaly detection task. On the other hand, the data profiling task only involves the steps of column sampling, constrained template generation, and pattern generation, as all data are assumed to be healthy for this task (i.e., $r_{cov}$ is constantly 1). 
Estimating the portion of healthy data (i.e., $r_{cov}$) is an essential process in generating patterns for anomaly detection, as healthy patterns should only be learned from healthy data. 
Since automated coverage rate estimation depends on template generation, pattern generation, and pattern selection, 
we introduce this step after explaining the other three steps. 

\subsection{Column Sampling}
\label{sec:column_sampling}

Typically, a data column \st{in a dataset}(e.g., a column about dates) only covers a subset of the entire data domain (all possible values of dates) \qql{that may come through the data pipeline}. Ideally, high-quality patterns should not only cover all the samples in the available data \qql{that can be seen during the pattern generation process,} 
\qql{but also be able to generalize to unseen future data}
\st{while being able to generalize to all samples in the same domain (including those unseen data)}
~\cite{song2021auto}. Therefore, we use a subset from the column \st{for pattern generation} \qql{as the pattern generation subset} 
and use the whole data column for \qql{generalizability evaluation and} pattern selection. 
To ensure the representativeness of the sample and the pattern generation efficiency, we use a sample size $N_{tr}$ which is calculated using a z-score with a confidence level of 95\% and an error margin of 5\%; $N_{tr}$ samples are randomly drawn from a column whose size is $N$ without replacement. 

\subsection{Constrained Template Generation}
\label{sec:template_clustering}

\begin{figure}[!t]
  \centering
  \includegraphics[width=0.8\columnwidth]{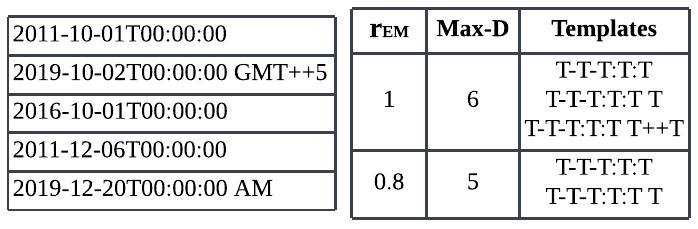}
  \vspace{-2ex}
  \caption{An example of 5 date time records and the templates generated under different exact matching rates ($r_{EM}$). T in templates represents tokens. Max-D stands for the maximum number of delimiters accepted under certain $r_{EM}$. } 
  \label{fig:datetime_example}
  \vspace{-2ex}
\end{figure}

Template generation without constraints may result in an over-abundance of templates.
Indeed, the basic structure of patterns can be captured by raw templates that contain \textbf{TOKEN}s and \textbf{delimiters}. 
We could finely split all records using all of their symbol strings (i.e., symbols containing no tokens within, such as ``++'' in Fig.~\ref{fig:datetime_example}) as delimiters. However, overly fine-grained splitting of the records may result in under-generalization, a problem in anomaly detection~\cite{he2021auto}. 
To prevent under-generalization, we establish an \textbf{exact matching rate} $r_{EM}$ to control the granularity of raw templates. 
$r_{EM}$ controls the portion of records that need to be \heng{fully split, thus determining the required maximum number of delimiters} \st{finely split and the maximum number of delimiters,}and the granularity of raw templates.
As illustrated in the right part of Fig~\ref{fig:datetime_example}, fully splitting all records (i.e., $r_{EM}$ = 1) 
results in the last two templates matching only one record. The scattered raw template distributions may cause the generated patterns to have low frequency, leading to a biased result in the pattern selection step (see Section~\ref{sec:cluster_for_pattern_selection}). 

Constrained template generation aims to convert records to raw templates under the constraint of $r_{EM}$. 
For the data profiling task, $r_{EM}$ is fixed to 1, as we aim to capture all template structures and do not require pattern selection. 
For data anomaly detection, we should only obtain templates for the exact matching of the healthy data. \heng{Hence, we set the exact matching rate with the same value as the estimated coverage rate (i.e., the estimated percentage of healthy data, see \ref{sec:coverage_rate_estimation}): $r_{EM}=r_{cov}$, 
with the intuition that all the healthy data would have exact template matching.} 
Fig~\ref{fig:datetime_example} is an example of a datetime column with five records 
from a sampled set ($N_{tr}=5$). 
First, we calculate the number of records to be fully split using $r_{EM}$. The number shall be calculated with $N_{tr}*r_{EM}$ and rounded to an integer. In our example, when setting $r_{EM}$ to 1, we should fully split all five samples; when $r_{EM}=0.8$, four samples are to be fully split. When $r_{EM}<1$, 
we determine the samples to be fully split using the number of delimiters they contain. According to the minimum description length principle~\cite{grunwald2007minimum}, we capture the minimum number of needed delimiters to fulfill the $r_{EM}$ constraint. The numbers of delimiters in the example are $\{4, 6, 4, 4, 5\}$. Hence, when $r_{EM}=0.8$, we accept five as the maximum number of acceptable delimiters, as it is the minimum number of delimiters required to fully split four samples. 

After determining the maximum 
number of acceptable delimiters, we generate raw templates by iterating through all the records. For records containing less or equal to the maximum number of delimiters, we fully split them and take all their symbol strings as delimiters. Otherwise, we stop splitting when the maximum value is reached. Under the setting of $r_{EM}=0.8$, the second record contains six delimiters and will be split as ``T-T-T:T:T T'' (i.e., matching the template of the last record). The portion ``T++T'' in the raw template is merged as a ``T'' after reaching the maximum delimiter number.

\subsection{Pattern Generation}
\label{sec:csp_pattern_generation}

\begin{figure*}[!t]
  \centering
  \includegraphics[width=\textwidth]{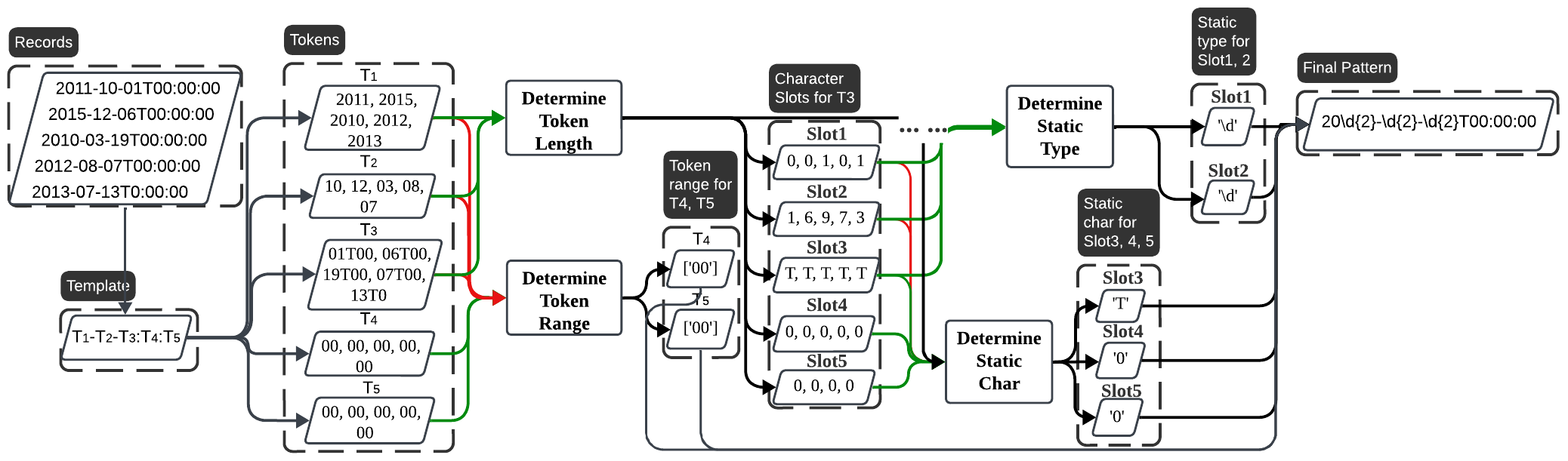}
  \vspace{-3ex}
  \caption{A date time example of using RIOLU generating patterns ($r_{cov}=0.8$). ``T'' stands for tokens. The green paths indicate a constraint has been found for the input (i.e., token or character slot) on the corresponding layer, while the red paths suggest that the input does not have a constraint on this layer. }
  \label{fig:generation_example}
  \vspace{-2ex}
\end{figure*}

Each record is matched to a raw template after the constrained template generation stage. We further elaborate on the content constraints and form patterns based on the templates. Raman \textit{et al.}~\cite{raman2001potter} use \textbf{token range} and \textbf{token length} to constrain the tokens. 
\qql{Instead of using the tokens, Ilyas \textit{et al.}~\cite{ilyas2018extracting} split the tokens into characters and discuss whether the character slot contains specific \textbf{static characters} or covers one or more \textbf{character type}.} \heng{Inspired by these approaches, we consider these four types of constraints for pattern generation: token range, token length, static character, and character type.} 


\st{The four constraint layers can be ordered from strict to loose: a token with a specific content range constraint (e.g., AM/PM) implicitly follows a length constraint, and the characters are fixed; if a token does not have a range constraint, its length may be limited (e.g., phone number). When a character slot has static characters, an implicit character type constraint is imposed (e.g., a static character range of 1,2,3 is only accepting digits). The character slot constraints can only be determined after the token constraints are fixed, as the character slot will be naturally invalid when they exceed the length constraint, and no constraint should be selected. On the other hand, if a token does not have a length constraint, we may fix character constraints on the maximum shared length (i.e., the smallest token length) and add suggestive constraints on where the slot position exceeds this length. }
\st{Based on this ordering, we use a water flow structure to select constraints and generate patterns. }
\qql{We use a waterfall structure to infer the constraints at the four layers, with an order from stricter to looser: 1) token range determination, 2) token length determination, 3) static character determination, and 4) static type determination. The intuition is simple: if a token has a stricter constraint (e.g., with a token range constraint: being either ``AM'' or ``PM''), then we do not need to infer the looser constraints (e.g., then token length or character-level constraints).}
We iterate records in the sampled set to collect all the contents on the four layers and select content constraints using $r_{cov}$ as a threshold. We use Fig~\ref{fig:generation_example} as a running example and assume $r_{cov}=r_{EM}=0.8$.  

\noindent\underline{\textbf{Content Collection.}}
Fig~\ref{fig:generation_example} illustrates five date time records that match with a template ``T-T-T:T:T''. Each token in the template is matched to a string in the records. For example, the first ``T'' in the first record corresponds to ``2011''. Following this method, we collect the contents of each token from the five examples, as shown in the ``Tokens'' frame. These contents are used for token range and token length constraint inference. 

The tokens can be further decomposed into character slots for static character and character type inference. In Fig~\ref{fig:generation_example}, we decompose the third token as an example. The character collection is similar to token content collection: we iterate through the tokens and fit their contents into the corresponding slot position, as shown in the ``Character Slots for T3'' frame. 


\noindent\underline{\textbf{Constraint Inference.}}
When selecting content constraints, Auto-Validate~\cite{song2021auto} aims to find patterns that capture most of the values under a small tolerance value $\theta$ (e.g., 1\%, 5\%) for anomalies in the training set. 
However, we argue that \st{the}\qql{using a constant} predefined $\theta$ is hardly precise\st{, }\qql{. Instead, the value should be adjusted based on the specific error rate of each dataset.} 
Therefore, we estimate the percentage of healthy values $r_{cov}$ to guide the constraint selection stage for the anomaly detection task. As mentioned in Sec~\ref{sec:overview}, for data profiling, the coverage rate is 1, given that the goal is to describe all data fully. 

\subsubsection{\textbf{Token Range Determination}}
\label{sec:token_range_determination}

\qql{Constant token contents (e.g., ``AM'' or ``PM'') are highly frequent~\cite{padhi2018flashprofile}.
} We use a two-class K-Means clustering technique to cluster the token values based on their frequency. We choose K-Means due to its efficiency and wide use for data clustering. 
The frequency range that can be reached by a token is: $[\frac{1}{N_{tr}}, 1]$.
To ensure that two clusters (i.e., the high and low-frequency cluster) can always be created, we manually insert 1 and $\frac{1}{N_{tr}}$ into the frequency list. The frequencies in the high-frequency cluster are then summed to compare with the coverage rate $r_{cov}$: 
if the sum is larger than $r_{cov}$, then the high-frequency values are making an adequate match, and thus a token range is found. Otherwise, we shall further determine whether the token has a length constraint. 
In our date time example, the template contains five tokens. As shown in Fig~\ref{fig:generation_example}, all the values for $T_4$ and $T_5$ share the value of ``00'': the token value ``00'' has a frequency of 1. Hence, a token range is assigned to these two tokens. Conversely, $T_1$, $T_2$, and $T_3$ do not have any token range since each token value has a frequency of $0.2$, and is clustered with $\frac{1}{6}\approx 0.167$ (i.e., the low-frequency cluster).

\subsubsection{\textbf{Token Length Determination}}
\label{sec:token_length_determination}
While a specific content range may not be applicable for a token, the token may have a limitation to the length. 
A static length $len$ for a token is determined if and only if most token contents have a fixed length; if the system fails to detect a static length for the current token, the minimum token length is captured as $len_{Min}$. In our running example in Fig~\ref{fig:generation_example}, we should detect token length constraints for $T_1$, $T_2$, and $T_3$. All the contents for $T_1$ and $T_2$ are in a fixed length, while $T_3$ contains four (80\%) 5-character contents and one (20\%) 4-character content. Under the setting of $r_{cov}=0.8$, we determine that $T_3$ also has a token length constraint since 80\% of the contents satisfy the constraint of having 5 character slots.  

\subsubsection{\textbf{Static Character Determination}}
\label{sec:static_character_determination}
We use K-Means clustering based on frequencies to determine static characters in the pattern, similar to the strategy for token range determination. The frequencies are calculated by counting the existence frequency 
for a character on a certain slot (e.g., for the first character slot in $T_3$, the frequency of ``0'' is 0.6 while the frequency for ``1'' is 0.4). We also insert 1 and $\frac{1}{N_{tr}}$ into the frequency list to ensure the split. 
For static length tokens (i.e., with $len$), we detect static character ranges for all $len$ characters; for tokens without static length constraints, we detect static character ranges for the first $len_{Min}$ characters.
We decompose $T_3$ in Fig~\ref{fig:generation_example} to demonstrate the approach. The constant length constraint is 5 for this token. According to the collected character in the ``Character Slots for $T_3$'' frame, the last three slots have static characters, but no static character is detected for the first two slots.

\subsubsection{\textbf{Static Type Determination}}
\label{sec:static_type_determination}
\qql{
Characters are usually categorized into upper/lower letters, digits, and symbols. }Static type is to be detected if and only if when a range of static characters (i.e., a stricter constraint) fails to be determined. 
For each character slot, the character types are ranked according to their frequencies from high to low, and the types are selected until their cumulative frequency exceeds $r_{cov}$ (i.e., the types cover the majority of the space). 
If all types appear in the majority of current character slot, the slot would have no constraint. 
Following the intuition in static character determination, for static length tokens, the type constraint applies to all $len$ characters; for tokens without static length constraints, we detect character types for the first $len_{Min}$ characters. 
For the example token $T_3$ in Fig~\ref{fig:generation_example}, since the first two slots in $T_3$ do not have static characters, we check their static character type. Both slots have a static type of digits, written as ``\textbackslash d''.

\subsubsection{\textbf{Pattern Composition}}
After constraints on the four layers are detected, we compose them into a pattern. For tokens with token range constraints, we directly replace the token using the range (e.g., the last two tokens in Fig~\ref{fig:generation_example} are replaced by ``00''). Otherwise, we write the regular expression for the token using detected constraints. Take $T_3$ as an example: using the ``\textbackslash d'' constraints for the first two character slots and the ``T00'' constraints for the last three slots, the regular expression for $T_3$ is ``\textbackslash d\{2\}T00''. Following this procedure, a pattern is generated as illustrated in the ``Final Pattern'' frame in Fig~\ref{fig:generation_example}.

\subsection{Pattern Selection}
\label{sec:cluster_for_pattern_selection}
\heng{A pool of candidate patterns is constructed after the pattern-generation step. However, not all patterns are considered as healthy patterns.}  
As shown in Fig~\ref{fig:methodology_overview}, we further perform pattern selection for the data anomaly detection task. 
Similar to Padhi \textit{et al.}~\cite{padhi2018flashprofile}, we consider patterns with low matching rates on the dataset as anomalies. 
For each pattern $p_i$, its matching rate (i.e., frequency) on the whole dataset is calculated through the number of records matched using regex. 
We then apply K-Means clustering to split the patterns into two clusters based on their matching rate 
automatically~\cite{lloyd1982least}: the \textbf{high-frequency} and \textbf{low-frequency} clusters. The clustering technique could avoid the domain-dependent threshold determination problem raised in Sec~\ref{sec:pattern_anomaly_detection}. 
We assume that 
at least one pattern shall be accepted for the corresponding column. 
Thus, we insert a low matching rate noise (i.e., $\frac{1}{N}$
) to ensure a low-frequency cluster is created. 
Patterns labeled as \textbf{high-frequency} are selected as healthy patterns, whereas those labeled as \textbf{low-frequency} are not further used. Finally, the healthy patterns are used to detect anomalous records in the column: records that do not match any healthy pattern are identified as anomalies. 


\subsection{Coverage Rate ($r_{cov}$) Estimation}
\label{sec:coverage_rate_estimation}

Coverage rate $r_{cov}$ is essential in template generation and pattern generation. 
Similar to column sampling (Sec~\ref{sec:column_sampling}), for the sake of generalizability and efficiency, we use a statistically representative sample (of size $N_{tr}$) of a data column to estimate the corresponding $r_{cov}$, either in a supervised or an unsupervised way.

\noindent\underline{\textbf{Guided-RIOLU: Supervised Estimation. }}
With human involved, it is feasible to notice pattern inconsistency in the sampled data and label them 
with domain knowledge. Users are required only to label whether a record in the sampled set (of size $N_{tr}$) is a pattern anomaly in the column. The labeled data can be used to calculate the portion of healthy data, which is the portion of data to be covered ($r_{cov}$). 

\begin{figure}[thb]
  \centering  \includegraphics[width=.8\columnwidth]{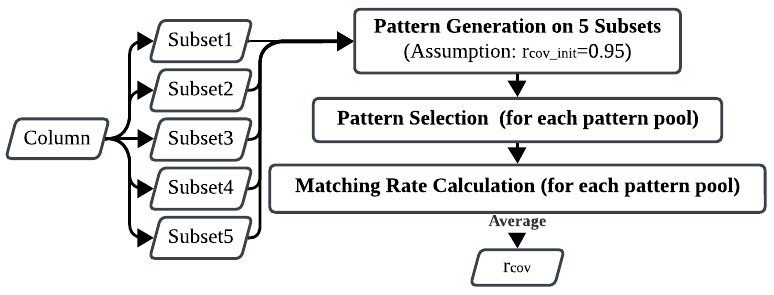}
  \caption{The coverage rate estimation process in unsupervised approach.} 
\label{fig:unsupervised_coverage_rate_estimation}
\end{figure}

\noindent\underline{\textbf{Auto-RIOLU: Unsupervised Estimation. }}Our intuition for pattern anomalies is that these anomalies cannot form large pattern clusters, similar to Padhi \textit{et al.}~\cite{padhi2018flashprofile}. 
Conversely, healthy patterns are highly frequent and robust enough to form a large cluster. As illustrated in Fig~\ref{fig:unsupervised_coverage_rate_estimation}, we apply a three-step approach to estimate the coverage rate. We first randomly draw five subsets \qql{(i.e., $N_{subset}=5$)} from the considered data column to ensure measurement stability, each with size $N_{tr} $. 
Then we generate five initial pattern pools from these five subsets with the assumption of \qql{$r_{cov\_init}=0.95$} \st{$r_{cov}=0.95$}, 
following the steps described in \ref{sec:template_clustering} and \ref{sec:csp_pattern_generation}. 
\heng{We choose $r_{cov\_init}=0.95$ as the initial overage rate since prior work shows that the anomaly data rates are typically less than 5\%~\cite{song2021auto}. We further evaluate the sensitivity of our approach for the settings of the $r_{cov\_init}$ value and the $N_{subset}$ value in \ref{sec:sensitivity_ablation}.}
In the second step, we select 
patterns with high matching rates from the five initial pattern pools through the pattern selection process (as described in \ref{sec:cluster_for_pattern_selection}). In the last step, each selected 
pattern pool is used to calculate the portion of matches they can create (i.e., matching rate) 
on the whole dataset. 
The matching rates provide us with an estimation of the portions of large 
clusters (i.e., potentially healthy 
clusters); 
noise patterns provide few matches 
and will be dropped in the pattern selection stage.
Using five subsets can statistically suppress the randomness of the evaluation process. Hence, we take the average matching rate as the estimated $r_{cov}$. 

\section{Experiments}
\label{sec:experiment}


This section describes our experiment design and results for evaluating RIOLU through our two research questions. 
RQ1 evaluates RIOLU's capacity to produce high-quality patterns, in a pattern-based data profiling setting. 
RQ2 evaluates RIOLU's capacity to detect
pattern anomalies from uncleaned data in an unsupervised or semi-supervised manner. 



\subsection{RQ1: How well can RIOLU profile data in different domains? }
\label{sec:data_profiling_evaluation}
Patterns extracted for data profiling should adequately and precisely describe the corresponding data domain. Further, it does not assume that the dataset is dirty in nature. Hence, the coverage rate to be reached in data profiling tasks shall be 1, 
and all the generated patterns should be accepted to describe the domain. Through the data profiling quality analysis, we could gain insight into RIOLU's pattern generation capability. 


\subsubsection{\textbf{Dataset}}
We choose a set of public datasets~\cite{dataprofilingdata} used to evaluate data profiling in prior work (FlashProfile~\cite{padhi2018flashprofile}
). 
To ensure generalizability, we consider the entire \textit{DOMAINS} data with 63 datasets with 45 to around 20k records, from a variety of domains (e.g., amino structure, software configuration, IP address
)\st{.}\qql{; the files, along with detailed domain information, are listed in our replication package.} 
Using these datasets allows us to have a fair comparison with prior work~\cite{padhi2018flashprofile}.

\subsubsection{\textbf{Methods Compared}}
\heng{We compare RIOLU with the state-of-the-art open-source tool for data profiling (FlashProfile~\cite{padhi2018flashprofile}), as well as ChatGPT which has recently shown promising results for regex generation~\cite{ChatGPT}.}
\begin{itemize}[leftmargin=*]
    \item\qql{\textbf{ChatGPT}~\cite{ChatGPT}. LLMs have shown their power in various tasks including regex inference~\cite{siddiq2024understanding}. We specified the pattern-based data profiling task to the GPT-3.5 Turbo API 
    and obtained the responses, as it was one of the most powerful and affordable at the experiment time. The regexes are manually extracted from the responses. We provide the prompt template and responses in the replication package. } 
    \item\textbf{FlashProfile~\cite{padhi2018flashprofile}}. FlashProfile is the state-of-the-art open-sourced tool for pattern-based data profiling. 
    FlashProfile clusters record using syntactic similarity and draw a pattern for each cluster. It significantly outperforms previous state-of-the-art tools, including MicrosoftSSDT~\cite{SSDTDocumentation} and Ataccama One~\cite{AtaccamaWebsite}. 
    \item\textbf{RIOLU}. 
    Data profiling does not require estimating the coverage rate, given that the patterns should cover all data. 
    To use RIOLU for data profiling, we set the coverage rate $r_{cov}$ to 1 and skip the pattern selection step. 
\end{itemize}


\subsubsection{\textbf{Evaluation and Metrics}} The goal of data profiling is to extract patterns that can fully and solely describe a domain. We followed the evaluation settings used in prior work~\cite{padhi2018flashprofile}. 20\% of the data from each dataset is randomly extracted as the \qql{pattern generation dataset}\st{training set (i.e., the set used for pattern generation)}. 
\qql{While prompting ChatGPT, we observed that, for 7 datasets containing many (i.e., $\sim$10k) records, or long content (e.g., protein structures), a random sample of 15\% of the data exceeded the token limit of ChatGPT. Thus, we only use 10\% of these datasets for training data.} 
The \textit{true positive rate} under this scenario is the portion of matches produced by the pattern pool in the remaining 80\% \qql{(or 90\% for 7 datasets with ChatGPT)} 
data from the same domain. An equal number of records are randomly drawn from other domains, and the \textit{false positive rate} is the matching rate on records belonging to the other domains. 
The intuition is that patterns drawn from a dataset shall describe all the records in this domain while not over-generalizing to other domains. We also calculate the \textit{precision}, \textit{recall}, and \textit{F1 score} based on these metrics.
\begin{table}[t]
\centering
\caption{The average performance of \qql{ChatGPT, }FlashProfile and RIOLU for data profiling (TP: average true positive rate; FP: average false positive rate; P: average precision; R: average recall; F1: average F1 score).}
\label{tab:data_profiling_result}
\begin{tabular}{l|l|l|l|l|l}
\hline
\textbf{}             & \textbf{TP} & \textbf{FP} & \textbf{P} & \textbf{R} & \textbf{F1} \\ \hline
\qql{\textbf{ChatGPT}}        & \qql{82.6\%}     & \qql{0.57\%}      & \qql{90.6\%}     & \qql{82.6\%}     & \qql{86.4\%}     \\ \hline
\textbf{FlashProfile} & 93.2\%      & 2.3\%       & 97.8\%     & 93.4\%     & 96.2\%     \\ \hline
\textbf{RIOLU}        & 95.1\%      & 0.16\%      & 99.4\%     & 95.1\%     & 97.2\%     \\ \hline

\end{tabular}
\label{tab:data-profiling-results}
\vspace{-2ex}
\end{table}

\subsubsection{\textbf{
Evaluation Result}}

\begin{figure}[!t]
  \centering
  \includegraphics[width=\columnwidth]{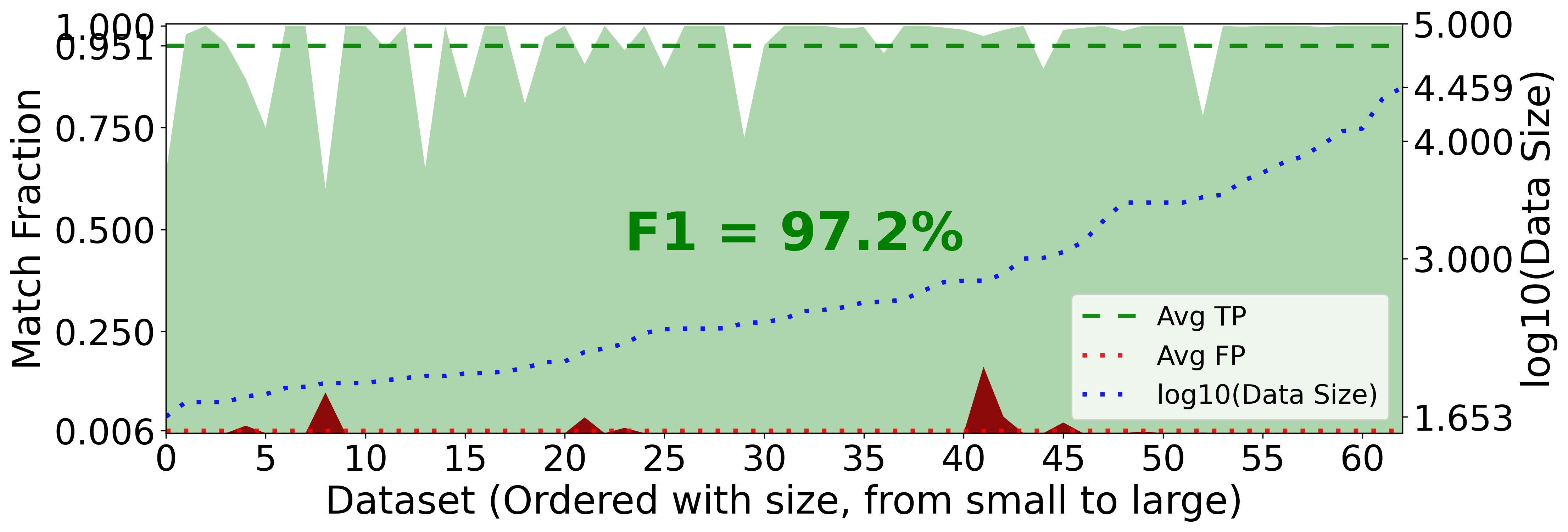}
  \vspace{-3ex}
  \caption{RIOLU's profiling quality on FlashProfile-DOMAINS dataset. The white peaks in the upper part denote the false negative rate, and the red peaks in the lower part indicate the false positive rate.
  }
  \label{fig:profiling_quality}
  \vspace{-3ex}
\end{figure}

Table~\ref{tab:data-profiling-results} shows the average performance of RIOLU and the baseline\qql{s}\st{(FlashProfile)} for data profiling. Fig~\ref{fig:profiling_quality} illustrates RIOLU's profiling quality in the same manner as FlashProfile~\cite{padhi2018flashprofile} does, showing the detailed performance for each individual dataset. As reported in \cite{padhi2018flashprofile}, FlashProfile has an average true positive rate of 93.2\%, an average false positive rate of 2.3\%, with an average F1 score reaching a high value of 96.2\%. 
According to Table~\ref{tab:data_profiling_result}, RIOLU 
pushed the accuracy higher with an average F1 score of 97.2\%. The average true positive rate obtained by RIOLU is around 95.1\%, while the false positive rate is 0.16\% (less than 10\% of that of FlashProfile). \qql{On the other hand, the overall performance of ChatGPT is worse than both FlashProfile and RIOLU: the average F1 score is only 86.4\%. In particular, ChatGPT's false positive rate is about 3.6 times that of RIOLU. Besides, ChatGPT generated one pattern that encountered issues during regex compiling (i.e., bad character range issue in a piece of regex ``[9-100]''), and 5 patterns that failed to match any record in the domain (e.g., ``[A-Fa-f0-9]\{24\}'' for code records with only 22 characters).} The results indicate that the patterns created using our raw templates and water flow constraint selection approaches can accurately capture pattern features for each domain and distinguish them from other domains. \qql{Moreover, we noticed that data-driven pattern profiling methods tend to be more accurate than ChatGPT. }



We also observe a general trend of improved performance of RIOLU when the data size increases 
in Fig~\ref{fig:profiling_quality}. This is because a small data size may fail to sample adequate supportive records as all the patterns may have a low frequency. Therefore, these patterns cannot be accurately captured. When the data size increases, the trend of highly frequent patterns becomes significant, and thus, RIOLU can learn more robust patterns. 


\begin{boxD}
    \noindent\underline{\textbf{Summary.}} RIOLU performs well on the data profiling task on 63 datasets from various domains. In particular, the false positive rate of RIOLU is less than 10\% of that of the state-of-the-art baseline (FlashProfile) \heng{and about 28\% of ChatGPT's}, indicating that Auto-RIOLU can describe the data more precisely and avoid over-generalization, which is essential for data anomaly detection. 
\end{boxD}

\subsection{RQ2: How precisely can RIOLU detect pattern anomalies?}
\label{sec:pattern_anomaly_detection_evaluation}
To answer this question, we evaluate RIOLU's capacity for detecting data pattern anomalies using \st{two}\qql{three} sets of data: open-sourced \qql{tabular} data\qql{, Java project method names,} and commercial data. 
We first compare Auto-RIOLU and Guided-RIOLU 
with the baselines 
on five open-sourced \qql{tabular}  datasets. 
\qql{We also explore the method naming patterns in popular Java projects and examine the inconsistencies using Auto-RIOLU.} On the commercial dataset, we predict patterns using Auto-RIOLU in different domains and ask the data analysis team in our industry partner CompanyX to validate the patterns. 

\noindent\underline{\textbf{Evaluation on Open-source Datasets. }}
\subsubsection{\textbf{Dataset}} To estimate the anomaly detection performance of RIOLU, we collected five publicly available datasets in different domains and with various sizes. These datasets contain multiple types of errors, including pattern violations and other types of data quality issues. 
\qql{We obtained the original data and their cleaned versions from previous data quality studies~\cite{mahdavi2019raha}~\cite{visengeriyeva2018metadata}~\cite{rekatsinas2017holoclean}~\cite{li2015truth}, and selected the columns containing pattern anomalies. }
Table~\ref{tab:dataset_info} shows the data size\qql{,}\st{and} error rate\qql{, and description of each dataset}.  
The data columns' domains include date, time, postal code, phone number, etc. 

\vspace{-2ex}
\begin{table}[h]
\centering
\caption{The information about the five public datasets. The size of each dataset is written as rows*columns.}
\label{tab:dataset_info}
\resizebox{\columnwidth}{!}{%
\begin{tabular}{l|l|l|l}
\hline
\textbf{Dataset}   & \textbf{Size} & \textbf{Error Rate} & \qql{\textbf{Description}} \\ \hline
\textbf{Hosp-1k} & 999*3   & 23.9\% & \multirow{3}{*}{\begin{tabular}[c]{@{}l@{}}\qql{Hospital data from US Department}\\\qql{of Health\& Human Services.} \end{tabular}} \\ \cline{1-3}
\textbf{Hosp-10k}  & 10000*3       & 23.2\%              &                                           \\ \cline{1-3}
\textbf{Hosp-100k} & 100000*3      & 23.7\%              &                                           \\ \hline
\textbf{Flights}   & 74066*6       & 52.5\%   &    \qql{Flight data from 38 airline-related websites.}        \\ \hline
\textbf{Movies}    & 7390*4        & 0.2\%    &  \qql{IMDb film data.}         \\ \hline
\end{tabular}%
}
\vspace{-2ex}
\end{table}
\begin{table*}[th]
\centering
\caption{The experiment results for quantitive analysis on five datasets. (P: precision; R: recall; F1: F1-Score)}
\resizebox{\textwidth}{!}{%
\begin{tabular}{c|ccc|ccc|ccc|ccc|ccc}
\hline
\multirow{2}{*}{\textbf{}} &
  \multicolumn{3}{c|}{\textbf{Hosp-1k}} &
  \multicolumn{3}{c|}{\textbf{Hosp-10k}} &
  \multicolumn{3}{c|}{\textbf{Hosp-100k}} &
  \multicolumn{3}{c|}{\textbf{Flights}} &
  \multicolumn{3}{c}{\textbf{Movies}} \\ \cline{2-16} 
 &
  \multicolumn{1}{c|}{P} &
  \multicolumn{1}{c|}{R} &
  F1 &
  \multicolumn{1}{c|}{P} &
  \multicolumn{1}{c|}{R} &
  F1 &
  \multicolumn{1}{c|}{P} &
  \multicolumn{1}{c|}{R} &
  F1 &
  \multicolumn{1}{c|}{P} &
  \multicolumn{1}{c|}{R} &
  F1 &
  \multicolumn{1}{c|}{P} &
  \multicolumn{1}{c|}{R} &
  F1 \\ \hline
  \textbf{\qql{ChatGPT}} &
  \multicolumn{1}{c|}{\qql{0.617}} &
  \multicolumn{1}{c|}{\qql{0.435}} &
  \qql{0.510} &
  \multicolumn{1}{c|}{\qql{0.667}} &
  \multicolumn{1}{c|}{\qql{0.280}} &
  \qql{0.385} &
  \multicolumn{1}{c|}{\qql{1.0}} &
  \multicolumn{1}{c|}{\qql{0.379}} &
  \qql{0.550} &
  \multicolumn{1}{c|}{\qql{0.958}} &
  \multicolumn{1}{c|}{\qql{0.594}} &
  \qql{0.733} &
  \multicolumn{1}{c|}{\qql{1.0}} &
  \multicolumn{1}{c|}{\qql{0.637}} &
  \qql{0.778} \\ \hline
  \textbf{FlashProfile} &
  \multicolumn{1}{c|}{0.333} &
  \multicolumn{1}{c|}{0.036} &
  0.065 &
  \multicolumn{1}{c|}{0.667} &
  \multicolumn{1}{c|}{0.070} &
  0.126 &
  \multicolumn{1}{c|}{0.366} &
  \multicolumn{1}{c|}{0.044} &
  0.079 &
  \multicolumn{1}{c|}{0.850} &
  \multicolumn{1}{c|}{0.641} &
  0.731 &
  \multicolumn{1}{c|}{1.0} &
  \multicolumn{1}{c|}{1.0} &
  \textbf{1.0} \\ \hline
  \textbf{Dynamic-FlashProfile} &
  \multicolumn{1}{c|}{0.333} &
  \multicolumn{1}{c|}{0.123} &
  0.181 &
  \multicolumn{1}{c|}{0.667} &
  \multicolumn{1}{c|}{0.170} &
  0.271 &
  \multicolumn{1}{c|}{0.667} &
  \multicolumn{1}{c|}{0.157} &
  0.254 &
  \multicolumn{1}{c|}{0.706} &
  \multicolumn{1}{c|}{0.658} &
  0.681 &
  \multicolumn{1}{c|}{1.0} &
  \multicolumn{1}{c|}{1.0} &
  \textbf{1.0} \\ \hline
\textbf{Auto-RIOLU} &
  \multicolumn{1}{c|}{0.712} &
  \multicolumn{1}{c|}{0.536} &
  0.611 &
  \multicolumn{1}{c|}{0.8} &
  \multicolumn{1}{c|}{0.35} &
  0.487 &
  \multicolumn{1}{c|}{0.8} &
  \multicolumn{1}{c|}{0.348} &
  0.485 &
  \multicolumn{1}{c|}{0.790} &
  \multicolumn{1}{c|}{0.692} &
  0.738 &
  \multicolumn{1}{c|}{1.0} &
  \multicolumn{1}{c|}{1.0} &
  \textbf{1.0} \\ \hline
\textbf{Guided-RIOLU} &
  \multicolumn{1}{c|}{0.66} &
  \multicolumn{1}{c|}{0.758} &
  \textbf{0.706} &
  \multicolumn{1}{c|}{1.0} &
  \multicolumn{1}{c|}{0.502} &
  \textbf{0.669} &
  \multicolumn{1}{c|}{1.000} &
  \multicolumn{1}{c|}{0.489} &
  \textbf{0.657} &
  \multicolumn{1}{c|}{0.824} &
  \multicolumn{1}{c|}{0.798} &
  \textbf{0.81} &
  \multicolumn{1}{c|}{1.0} &
  \multicolumn{1}{c|}{1.0} &
  \textbf{1.0} \\ \hline
\end{tabular}%
}
\label{tab:quantitive_result}
\vspace{-3ex}
\end{table*}

\subsubsection{\textbf{Methods Compared}} \heng{We compare Auto-RIOLU and Guided-RIOLU 
with ChatGPT and two versions of FlashProfile, the state-of-the-art open-sourced approach.}
\begin{itemize}[leftmargin=*]
    \item\qql{\textbf{ChatGPT}. Constrained by the token limit of ChatGPT, we randomly sample 3000 records for each column and present them to GPT-3.5 Turbo API for regex inference. For certain columns with a larger number of tokens, we fed 2500 records or 2000 records to meet the token limit. 
    The inferred regexes are then used for anomaly detection on the entire dataset. The sampled dataset and zero-shot prompt template can be found in our replication package. }
    \item\textbf{FlashProfile}~\cite{padhi2018flashprofile}. 
    As mentioned in \ref{sec:data_profiling_evaluation}, FlashProfile clusters data with different patterns and infers regular expressions for each cluster. Pattern anomalies can be found using a frequency threshold: patterns that are rare tend to be anomalies. We follow the suggestion provided in the instructions~\cite{proseinspector} 
    and set the threshold to 0.01. 
    \item\textbf{D-FlashProfile (Dynamic-FlashProfile)}. 
    D-FlashProfile is a combination of FlashProfile with our automated 
    pattern selection step (see Sec~\ref{sec:cluster_for_pattern_selection}). 
    Based on our pattern selection step, we avoid a fixed threshold but dynamically select the patterns based on their frequencies. The cluster with a higher frequency is accepted as the healthy pattern pool. 
    \item\textbf{Auto-RIOLU}. Auto-RIOLU learns and selects patterns using automated coverage ($r_{cov}$) estimation on unlabeled data (Sec~\ref{sec:coverage_rate_estimation}). 
    \item\textbf{Guided-RIOLU}. Guided-RIOLU estimates $r_{cov}$ according to the error rate of a labelled subset of the data (Sec~\ref{sec:coverage_rate_estimation}). 
\end{itemize}

We also considered XSystem~\cite{ilyas2018extracting} for evaluation. However, given that the system heavily relies on determining the number of branches (i.e., the number of different patterns) which requires domain expertise
, we failed to obtain reliable patterns. 
Thus, we excluded the tool from the comparison. 

\subsubsection{\textbf{Evaluation and Metrics}} As anomalies have the nature of being rare in the wild, the healthy 
and anomaly records are often imbalanced: an approach that constructs a broad pattern and predicts only the majority class (e.g., predicting every record as "normal" using ``.*") can achieve high accuracy without actually detecting any anomalies, which can give a biased impression of the performance. Thus, we do not evaluate the accuracy. Anomaly detection aims to identify all the anomalies while not misidentifying any healthy data. Hence, we leverage \textit{precision}, \textit{recall}, and \textit{F1 score} for the evaluation. Precision indicates the percentage of predicted anomalies that are actually anomalies; recall indicates the percentage of anomalies that are correctly predicted; the F1 score is the harmonic average of precision and recall.



\subsubsection{\textbf{Quantitative Evaluation Result}}

\begin{table}[]
\centering
\caption{\qql{The inference time statistics on the datasets (by second) of four compared tools. }}
\label{tab:time_comparison}
\resizebox{\columnwidth}{!}{%
\begin{tabular}{l|l|l|l|l}
\hline
\textbf{Time(s)} & \textbf{ChatGPT}   & \textbf{FlashProfile} & \textbf{Auto-RIOLU} & \textbf{Guided-RIOLU} \\ \hline
{[}Min, Max{]}   & {[}10.95, 27.44{]} & {[}17.08, 406.26{]}   & {[}0.34, 46.43{]}   & {[}0.27, 7.91{]}      \\ \hline
Average          & 13.72              & 111.16                & 12.35               & 2.63                  \\ \hline
\end{tabular}%
}
\vspace{-3ex}
\end{table}

\qql{We ran the experiment on a Mac desktop with an 8-core CPU (M2 chip) and 16G memory. The time used for pattern inference is reported in Table~\ref{tab:time_comparison}. 
The response time of ChatGPT was captured as the inference time. 
We did not include the execution time of using the generated patterns with regex matching to detect anomalies, since all tools use the same piece of regex matching code for the detection and the time consumption is neglectable. 
We did not include D-FlashProfile's inference time because this approach combines FlashProfile and our pattern selection component; the time consumption is larger than FlashProfile. Since the ground truths have been obtained from previous studies, the labeling time is excluded for Guided-RIOLU. Both versions of RIOLU require less inference time on average (Auto-RIOLU requires 10.0\% less time than ChatGPT, and Guided-RIOLU requires 80.8\% less). }

The performance results generated by the \st{four}\qql{five} tools compared are shown in Table~\ref{tab:quantitive_result}. To suppress the bias caused by random sampling, the results for both versions of RIOLU are the average on five detection runs. According to the table, both versions of RIOLU outperform \st{the baselines on four out of five datasets. On the $Movies$ dataset, all tools reached an F1 score of 1.0.}\qql{all the baselines on four out of five datasets. ChatGPT obtained high precisions and showed great potential in anomaly detection; on the \textit{Hosp-100k} dataset, ChatGPT detected more true positives in the state code domain and obtained a higher F1 score than Auto-RIOLU.} 

We noticed that Guided-RIOLU can boost performance on four 
datasets using a set of labeled data samples, achieving the best performance on all datasets. On four datasets, Guided-RIOLU obtained up to 37.4\% improvement of F1 over Auto-RIOLU. The sample size to be labeled is less than 0.4\% of the whole dataset. With the labeled small subset, RIOLU can estimate the coverage rate more accurately. 

It is also shown in Table~\ref{tab:quantitive_result} that combining our pattern selection step with FlashProfile (i.e., Dynamic-FlashProfile) enhanced its F1 score on three out of the five datasets: the automated selection of patterns can increase the quality of pattern pools on the three hospital datasets, proving the efficiency of the selection step. However, we noticed a performance drop in the \textit{Flights} dataset. The ground truth of the \textit{Flights} dataset contains scattered patterns (i.e., records with different patterns are accepted as ground truths). Since all the patterns do not have high coverage (i.e.,  with coverage around 10\%), they cannot be well-clustered based on their frequency.


\vspace{-1ex}
\begin{table}[!ht]
\centering
\caption{Patterns generated by \qql{ChatGPT}, FlashProfile, Dynamic-FlashProfile(D-FlashProfile), Auto-RIOLU, and Guided-RIOLU for four example data domains.}
\label{tab:qualitative_results}
\resizebox{\columnwidth}{!}{%
\begin{tabular}{l|l|l|l|l}
\hline
\textbf{} &
  \textbf{State Code} &
  \textbf{Zip} &
  \textbf{Phone} &
  \textbf{Duration} \\ \hline
  \textbf{\qql{ChatGPT}} &
  \begin{tabular}[c]{@{}l@{}}{[}A-Z{]}(?!{[}A-Z{]})\\{[}A-Z{]}\{2\}\end{tabular} &
  \begin{tabular}[c]{@{}l@{}}\textbackslash d\{5\}\\\textbackslash d\{5\}(?:-\textbackslash d\{4\})?\end{tabular}
  &
  \textbackslash d\{10\} &
  \begin{tabular}[c]{@{}l@{}}(\textbackslash d+)\textbackslash s*hr\textbackslash.\textbackslash s*\\(\textbackslash d+)\textbackslash s*min\end{tabular} \\ \hline
\textbf{FlashProfile} &
  {[}a-zA-Z{]}+ &
  \begin{tabular}[c]{@{}l@{}}{[}0-9{]}\{5\}\\ {[}0-9{]}\{4\}\\ {[}0-9{]}\{3\}\end{tabular} &
  \begin{tabular}[c]{@{}l@{}}{[}0-9{]}+\\ {[}0-9{]}+{[}a-z{]}{[}0-9{]}+\\ {[}0-9{]}\{10\}\textbackslash{}*\textbackslash{}*\textbackslash{}*\\ {[}a-z{]}{[}0-9{]}\{10\}\end{tabular} &
  \begin{tabular}[c]{@{}l@{}}{[}0-9{]}+\textbackslash\ min\\\textbackslash \ {[}0-9{]}\{4\}\textbackslash\ {[}A-Z{]}{[}a-z{]}+"\end{tabular} \\ \hline
\textbf{D-FlashProfile} &
  {[}a-zA-Z{]}+ &
  \begin{tabular}[c]{@{}l@{}}{[}0-9{]}\{5\}\end{tabular} &
  \begin{tabular}[c]{@{}l@{}}{[}0-9{]}+\end{tabular} &
  \begin{tabular}[c]{@{}l@{}}{[}0-9{]}+\textbackslash\ min\end{tabular} \\ \hline
\textbf{Auto-RIOLU} &
  {[}A-Z{]}\{2\}{[}A-Z{]}* &
  \textbackslash{}d\{5\} &
  \textbackslash{}d\{10\}&
  \textbackslash{}d\{2\}\textbackslash{}d*\textbackslash min \\ \hline
\textbf{Guided-RIOLU} &
  {[}A-Z{]}\{2\} &
  \textbackslash{}d\{5\} &
  \textbackslash{}d\{10\} &
  \textbackslash{}d\{2\}\textbackslash{}d*\textbackslash min \\ \hline
\end{tabular}%
}
\end{table}
\subsubsection{\textbf{Qualitative Evaluation Result}}
We extracted the patterns generated by \qql{ChatGPT,}  FlashProfile, D-FlashProfile, Auto-RIOLU, and Guided-RIOLU in four domain fields to analyze the results qualitatively. The first three domains are state code, zip, and phone number from the \textit{Hosp-100k} dataset and the duration domain from the \textit{Movies} dataset. We examine the quality of the patterns using domain knowledge gained through observing the records. According to Table~\ref{tab:qualitative_results}, Guided-RIOLU can generate and select correct patterns corresponding to prior knowledge for all four domains. Given that the coverage rate is automatically estimated, Auto-RIOLU provides a false positive pattern (i.e., strings containing more than two upper letters are accepted as state code). 

\qql{ChatGPT has external domain knowledge from pre-training. 
Hence, it inferred and selected the patterns based on both the provided column and its training data. For state code, it included a false positive regex matching any single uppercase letter that is not immediately followed by another uppercase letter. For zip code, ChatGPT accepted another format (i.e., \textbackslash d\{5\}(?:-\textbackslash d\{4\})) apart from five digits. The duration in the \textit{Movies} dataset should use minutes as units instead of hours. However, ChatGPT accepted the records with hours as a unit based on its external knowledge of time calculation, leading to more false negatives in anomaly detection.}

It can be observed that FlashProfile accepts several sub-optimal patterns with the default frequency threshold (e.g.,  phone numbers shall only contain digits; however, it provides us with three patterns, including either letters or symbols). With our pattern selection module attached, the patterns learned by Dynamic-FlashProfile are more precise. Although the length constraints cannot be determined accurately due to the pre-clustering technique of FlashProfile, Dynamic-FlashProfile eliminated patterns with significantly wrong types; for example, the three phone number patterns containing letters or symbols are dropped after clustering. 

The qualitative result suggests that although Auto-RIOLU can learn constraints on character types and length constraints, the learning quality is still to be improved with a more accurate estimation of the coverage rate. The estimation can be more accurate with humans involved to label a small subset (e.g., with around 350 samples).
\qql{We also found that RIOLU's accuracy can be limited when patterns are scattered (i.e., when a data column has many low-frequency healthy patterns). 
In such cases, RIOLU may fail to distinguish anomalies from low-frequency healthy patterns (Also see Sec~\ref{sec:validity_threat}). }

\begin{table*}
    
\centering
\caption{The method names consistent rate (i.e., the portion of naming matched with the inferred pattern), the number of anomalies detected, and the false positive rate. }
\label{tab:java_naming}
\resizebox{\textwidth}{!}{%
\begin{tabular}{l|l|l|l|l|l|l|l|l|l|l|l}
\hline
\textbf{Project Name} &
  \textbf{presto} &
  \textbf{wildfly} &
  \textbf{libgdx} &
  \textbf{intellij-community} &
  \textbf{gradle} &
  \textbf{liferay-portal} &
  \textbf{spring-framework} &
  \textbf{cassandra} &
  \textbf{hibernate-orm} &
  \textbf{hadoop-common} &
  \textbf{elasticsearch} \\ \hline
\textbf{Consistent Rate} & 0.998 & 0.980 & 0.963 & 0.992 & 0.999 & 0.813 & 0.982 & 0.949 & 0.996 & 0.989 & 0.980 \\ \hline
\textbf{\# Detected Anomalies}         & 4   & 361 & 619 & 1488  & 17 & 379 & 600  & 1029 & 96 & 390 & 435 \\ \hline
\textbf{FP Rate}         & 0.5   & 0.100 & 0.018 & 0     & 0.294 & 0.292 & 0.36  & 0.052 & 0.042 & 0.167 & 0.290 \\ \hline
\end{tabular}%
}
\vspace{-2ex}
\end{table*}


\noindent\underline{\qql{\textbf{Detection of Naming Inconsistency in Java Projects.}}}

\qql{Identifier (e.g., variable or method) naming has been widely recognized as an essential software engineering task~\cite{mcconnell1993code}~\cite{beck2007implementation}~\cite{martin2009clean}. Indeed, inappropriate naming can cause problems in developer comprehension~\cite{takang1996effects} and even code quality degradation~\cite{butler2009relating}. 
Oracle~\cite{oracle} suggests that Java method names should start with an upper letter and with the first letter of each internal word capitalized. }

\heng{We want to evaluate the ability of Auto-RIOLU to infer project-wise naming conventions and detect anomalies (i.e., inconsistencies) automatically, thus we do not specify any naming convention with Auto-RIOLU.}
\qql{To this end, we collected method names in the 11 top Java projects from the CodeAttention training set~\cite{allamanis2016convolutional}.
We ran Auto-RIOLU with default settings, obtained the inferred patterns, and detected anomalies. For all projects, our inferred healthy patterns match the Oracle Java method naming convention. 
\heng{In other words, all the detected normal names match the standard naming convention (i.e., no false negatives). Thus, we only manually analyze the detected anomalies, to understand the categories of anomalies and potential false positives. }
Given that the \textit{liferay-portal} project contains more than 20k detected anomalies, we randomly sampled 379 anomaly examples (i.e., 95\% confidence level with 5\% error rate) for labeling. Table~\ref{tab:java_naming} shows the detection and labeling results. The labels can be found in our replication package. }

\qql{As shown in Table~\ref{tab:java_naming}, 9 out of the 11 projects have a consistent naming rate higher than 95\%. There are two categories of naming anomalies (true positives): 1) method names containing symbols such as ``\_'' and ``\$'', and 2) method names starting with an uppercase letter or a digit.  
For the project with the lowest consistent naming rate (i.e., liferay-portal), 70.6\% 
of the detected anomalies contain symbols. }\qql{The detailed statistics can be found in our replication package. }

\heng{For 9 out of the 11 projects, Auto-RIOLU's false positive rates are relatively low, ranging from 0 to 29.2\%. For one project (liferay-portal), the false positive rate in the detected anomalies is 50\%
, which may be caused by its low anomaly rate: there are only 4 detected anomalies. 
Among these projects, the false positives are mainly caused by short names (e.g., ``go'', ``f''): Auto-RIOLU tends to treat such short names as anomalies since typical Java method names tend to have more than three characters.
}

\noindent\underline{\textbf{Evaluation on CompanyX database.}}

The data pattern inconsistency would block the automatic software pipeline and lead to software quality problems. Our industrial partner \textit{CompanyX} \heng{uses a large data warehouse to manage data from various sources and domains; the data are produced by some software (sub)systems (e.g., web applications) and consumed by other (sub)systems (e.g., data analytics pipelines). }\textit{CompanyX} faces the problem of data patterns not being aligned. \qql{The data pattern inconsistency would block the automatic software pipeline and lead to software quality problems.} \st{Given}\qql{However, given} the large volume of data, it is a great challenge to design all the regular expressions for each data domain manually. We \st{try to estimate the manual effort our tool can reduce for a real business, using}\qql{deployed RIOLU on the data warehouse management platform (Databricks) and evaluated its ability in generating patterns and detecting potential anomalies. We performed our evaluation using} three representative large tables (i.e., tables containing over \heng{20} columns and 500k records). Before executing the tool, we cooperated with the data analysis team to determine columns in target tables that require pattern validation. According to our discussion, the columns to be checked cover various domains, including date, ID, emails, province code, etc. The error rate of each column remains unknown before our detection. 
\begin{figure}[]
  \centering  \includegraphics[width=\columnwidth]{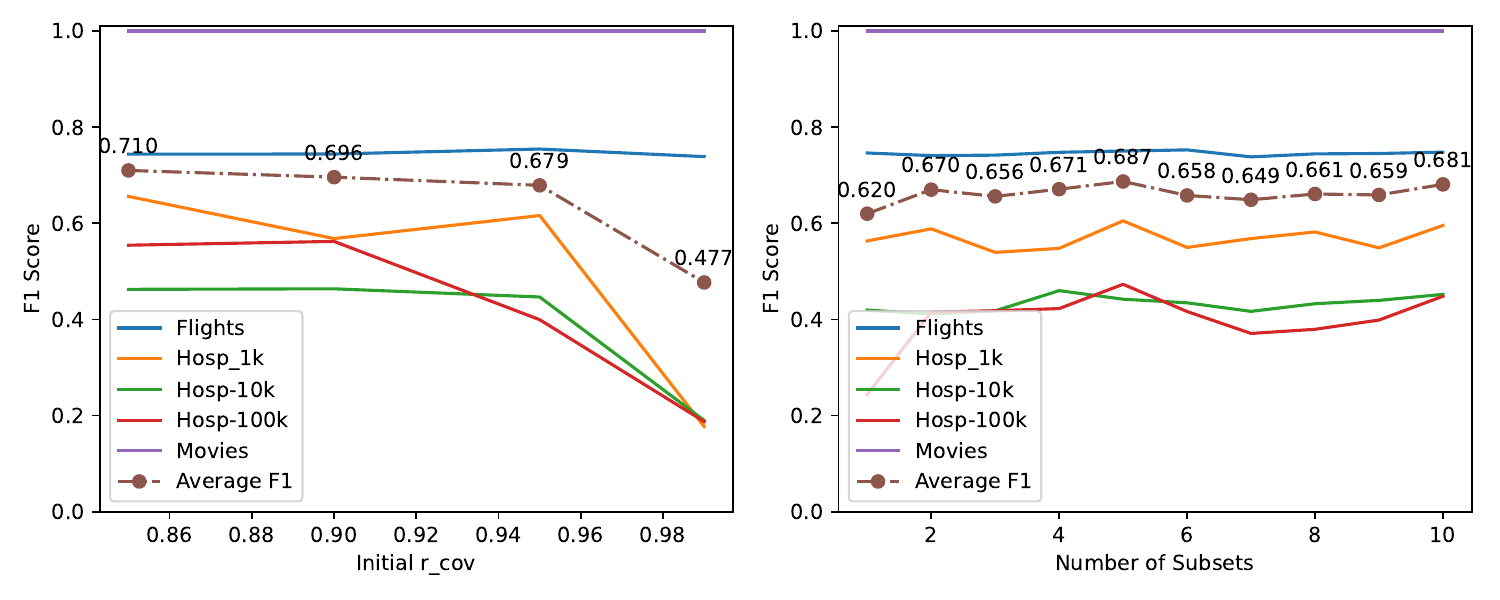}
  \vspace{-3ex}
  \caption{The impact of choosing different $r_{{cov}\_{init}}$ values (0.85, 0.9, 0.95, and 0.99) and $N_{subset}$ (range from 1 to 10) for Auto-RIOLU.}
  \label{fig:sensitivity_analysis}
  \vspace{-3ex}
\end{figure}
\heng{We then used Auto-RIOLU with the default setting to automatically generate the data patterns and detect potential anomalies for these columns}.
The team manually verifies the validity of each generated pattern by inspecting records in the corresponding columns. Auto-RIOLU can generate robust patterns for most columns and detect inconsistent data records (i.e., anomalies). \qql{For instance, based on Auto-RIOLU's results, we noticed that 0.15\% of liability codes are ``Unknown'', which may need the production team's attention; account IDs should contain a fixed number of digits, but 1.5\% of the records have abnormal patterns and require further investigation. }We observed that only email pattern generation had been complex, according to the heterogeneity of email addresses (e.g., john.doe@school.edu and johndoe@school.com can both be valid addresses, but they belong to different templates due to the symbol ``.'' existing in the first email address). Future work is needed to improve RIOLU for data with such such heterogeneous patterns.

\subsection{Ablation Study \& Sensitivity Analysis} 
\label{sec:sensitivity_ablation}

\qql{Several design choices and parameters may affect the performance of RIOLU. Thus, we carry out an ablation study and a sensitivity analysis to discuss their impacts.} 

\qql{\textit{Ablation study.} As described in Sec~\ref{sec:approach}, 
RIOLU takes five steps to automatically infer patterns and detect pattern anomalies: column sampling, coverage rate ($r_{cov}$) estimation, constrained template generation, pattern generation, and pattern selection. 
To understand the impacts of these steps and the design decisions therewithin, we carried out an ablation study on Auto-RIOLU for pattern anomaly detection using the 5 public datasets in RQ2. 
The ablation study considers five variants of Auto-RIOLU: 1) In the column sampling step, sampling 20\% of a column's records (similar to FlashProfile~\cite{padhi2018flashprofile}) instead of using a statistically representative sample; 2) removing the $r_{cov}$ estimation step and using a static default $r_{cov}$ of 0.95; 3) in the constrained template generation step, removing the constraint, generating templates that give exact match for every record (i.e., $r_{EM}$=1); 4) using a static pattern selection threshold (0.01 suggested by the online document of FlashProfile) 
instead of K-Means clustering; and 5) selecting all generated patterns (i.e., no pattern selection).
}

\qql{Table~\ref{tab:ablation_auto_riolu} presents the results of our ablation study. Overall, the original Auto-RIOLU has the highest and most stable performance. We observed that removing or modifying the pattern selection step (the ``No P.S.'' and ``Static P.S. Threshold'' rows in Table~\ref{tab:ablation_auto_riolu}) leads to the most significant performance drop, especially when the error rate is too high (e.g., the \textit{Flights} dataset, with F1 from 0.738 to 0.135, an 81.7\% drop): without proper pattern selection, anti-patterns may be selected, thus $r_{cov}$ cannot be accurately estimated. 
Besides, subsampling 20\% of the records might lead to unstable results, as a fixed-rate sampling may lead to under- or over-generalization of the original population. 
In addition, fixing $r_{cov}$ to an initial default value such as 0.95 (i.e., without automated $r_{cov}$ estimation) will allow RIOLU to over-optimistically assume that a fixed percentage (e.g., 95\%) of the records are healthy, leading to sub-optimal results; 
constantly setting $r_{EM}$ to 1 can over-generate trivial patterns and reduce their generalizability. 
}


\qql{\textit{Sensitivity analysis.} As introduced in Sec~\ref{sec:coverage_rate_estimation}, the $r_{cov}$ estimation relies on the $r_{{cov}\_{init}}$ and the number of subsets sampled from the column ($N_{subset}$). To validate the impact of our default choice (i.e., $r_{cov\_init}$=0.95, $N_{subset}$=5), we carried out a sensitivity analysis. We vary  $r_{cov}$ from 0.85 to 0.99 in increments of 0.05 (except the last increment to 0.99); and $N_{subset}$ from 1 to 10 in increments of 1. }

\qql{
We observe that the performance of Auto-RIOLU is relatively stable across different parameter settings in Fig.~\ref{fig:sensitivity_analysis}. The average F1 score across the datasets only changes 3.1\% when we vary the $r_{{cov}\_{init}}$ from 0.85 to 0.95; further increasing the $r_{{cov}\_{init}}$ (i.e., an initial too-high assumption of the health rate) would notably decrease the performance. 
The optimal choice of $r_{{cov}\_{init}}$ varies across datasets, given that their error rates are drastically different. 
Nevertheless, we recommend the use of the default parameter setting ($r_{{cov}\_{init}} = 0.95$) as it provides generalizable and near-optimal performance for most of the datasets, including those not used when building our tool. }
\qql{While changing the number of subset samples ($N_{subset}$), Auto-RIOLU also shows a relatively stable performance. We observed a peak of average F1 score when setting $N_{subset}$ to 5: a smaller number may cause the samples being less representative, whereas a larger number would decrease the efficiency and may cause too many overlaps among the samples. 
Hence, we selected 5 as the default $N_{subset}$ value. }


\begin{table}[]
\centering
\caption{The ablation study of Auto-RIOLU. The bold values indicate the highest, and the italic values show the lowest F1 scores for each dataset, respectively. ``Pattern selection'' is abbreviated as ``P.S.''. }
\resizebox{\columnwidth}{!}{%
\begin{tabular}{c|c|c|c|c|c}
\hline
\multirow{1}{*}{\textbf{}} &
  \multicolumn{1}{c|}{\textbf{Hosp-1k}} &
  \multicolumn{1}{c|}{\textbf{Hosp-10k}} &
  \multicolumn{1}{c|}{\textbf{Hosp-100k}} &
  \multicolumn{1}{c|}{\textbf{Flights}} &
  \multicolumn{1}{c}{\textbf{Movies}}\\\hline
  \textbf{Original} &
  {\textbf{0.611}} &
  {\textbf{0.487}} &
  {\textbf{0.485}} &
  0.738 &
  {\textbf{1.0}} \\ \hline
  \textbf{20\% Column Sampling} &
  {\textit{0.436}} &
  {\textit{0.362}} &
  0.462 &
  {\textbf{0.772}} &
  {\textbf{1.0}} \\ \hline
  \textbf{Static $r_{cov}$=0.95} &
  0.509 &
  0.440 &
  0.408 &
  0.569 &
  {\textbf{1.0}} \\ \hline
\textbf{Static $r_{EM}$=1} &
  0.537 &
  0.445 &
  0.323 &
  0.767 &
  {\textbf{1.0}} \\ \hline
\textbf{Static P.S. Threshold} &
  0.521 &
  0.428 &
  {\textit{0.296}} &
  0.147 &
  0.468 \\ \hline
  \textbf{No P.S.} &
  0.541 &
  0.435 &
  0.415 &
  {\textit{0.135}} &
  {\textit{0.360}} \\ \hline
\end{tabular}%
}
\label{tab:ablation_auto_riolu}
\vspace{-2ex}
\end{table}

\begin{boxD}
    \noindent\underline{\textbf{Summary.}} 
    The fully automated version of RIOLU (Auto-RIOLU) outperforms the baselines \qql{on four out of five public datasets and achieves better efficiency.} Guided-RIOLU further improves the performance (with up to 37.4\% improvement of F1 over Auto-RIOLU) with the guidance of a small set of labeled data. 
    \heng{Our extended evaluation on Java naming inconsistency detection and in an industry setting further demonstrates RIOLU's generalizability in detecting data pattern anomalies, hence contributing to the overall quality of software that relies on data.}
\end{boxD}

\section{Related Works}
\label{sec:related_works}

Data quality has incited tremendous interest in the software engineering (e.g., \cite{davoudian2020big, lwakatare2021experiences, foidl2022data, shome2022data}) and data engineering (e.g., \cite{ilyas2018extracting, song2021auto, breck2019data, Schelter2018}) communities. In particular, pattern violation is a major type of data quality issue as it directly impacts the quality of downstream software~\cite{abedjan2016detecting,visengeriyeva2018metadata,mahdavi2019raha,lwakatare2021experiences}. Below, we discuss prior work related to pattern-based data profiling, \qql{pattern} anomaly detection, \heng{and work on language mining which shares some similarity with pattern inference.}


\noindent\underline{\textbf{Pattern-Based Data Profiling.}} The goal of pattern-based data profiling is to describe data with a set of patterns, to help understand the data~\cite{padhi2018flashprofile}. 
Such comprehension of data is important for improving and maintaining data 
for software development and operations~\cite{lwakatare2021experiences}. 
Pattern-based data profiling tools use syntactic structures to describe data. Microsoft’s SQL Server Data Tools~\cite{SSDTDocumentation} 
pioneered learning descriptive regular expressions, but they lack extensibility and comprehensiveness. 
Ataccama One~\cite{AtaccamaWebsite} is a commercial offering that profiles the data using a set of base patterns. Potter’s Wheel~\cite{raman2001potter} describes the data domain using the most frequent pattern. FlashProfile~\cite{padhi2018flashprofile} clusters records according to syntactic similarity and assigns a pattern to each cluster. 
However, we observe that these approaches (e.g., FlashProfile) are not sensitive to subtle data noises: they could lead to false positives caused by over-generalization. 
Hence, we propose a novel approach to generate precise and general patterns, even when noises exist in the data (RIOLU can automatically distinguish healthy patterns from noises). 

\noindent\underline{\textbf{Pattern Anomaly Detection.}}
Pattern-based anomaly detection leverages manually designed or automatically inferred patterns to detect data anomalies~\cite{song2021auto}.
TFDV~\cite{breck2019data} and Deequ~\cite{Schelter2018} require manual design of regular expressions as patterns. These approaches are not effective when users lack knowledge of their data, and would be time-consuming for a large volume of data. 
Data-Scanner-4C~\cite{yu2023human} considers a human-in-the-loop method and lets users select healthy examples for pattern inference and inconsistency identification. However, example selection can still be labor-consuming since users must inspect the dataset for multiple loops.
To save manual effort, some studies focus on automated pattern inference. Potter’s wheel~\cite{raman2001potter} leverages minimum description length to extract patterns for each column; XSystem~\cite{ilyas2018extracting} infers pattern for every column with a branch-merge technique; and FlashProfile~\cite{padhi2018flashprofile} first clusters records by domain using syntactic similarity, then assign patterns to each cluster.  
Auto-Detect~\cite{huang2018auto} and Auto-Validate~\cite{song2021auto} both consider pattern inference with multiple columns by leveraging knowledge from similar columns, aiming to infer precise and generalizable patterns and suppress the false positive rate for anomaly detection. 
However, we noticed that these previous 
approaches require domain-specific thresholds to identify pattern anomalies, posing difficulty for users to use them for different data domains. Therefore, in this work, we propose a fully automated, unsupervised, and auto-parameterized approach for pattern-bassed anomaly detection.

\noindent\qql{\underline{\textbf{Language Mining.}} The task of pattern inference is related to the field of language mining and general grammar induction. For example, the L* algorithm~\cite{angluin1987learning} is an impacting method for inferring language using labeled samples.
Recent works such as Glade~\cite{bastani2017synthesizing}, Mimid~\cite{gopinath2020mining}, and Arvada~\cite{kulkarni2021learning} further enhanced the generalizability and speed of context-free grammar inference with positive examples; there is also specific research on grammar inference for Ad-Hoc parsers~\cite{schroder2022grammars}. However, these works do not explicitly support the detection of anomalies and do not offer an unsupervised variant. Users have to determine whether an input is a positive sample before feeding it into the algorithm. 
Moreover, although context-free grammars are expressive, the rules are more difficult for non-expert users to understand, validate, and modify. Conversely, Regexes are easier to write and expressive to validate tabular data with patterns (e.g., DateTime, URL), thus are used by IT companies such as Microsoft~\cite{SSDTDocumentation, padhi2018flashprofile} and Amazon~\cite{schelter2018automating}. }




\section{Threats to Validity}
\label{sec:validity_threat}

\noindent\underline{\textbf{Construction Validity. }}
Given the potential noise in the data and randomness in our approach (e.g., from column sampling), the 
pattern inference results and the performance of RIOLU may fluctuate. Thus, we ran the tests on RIOLU five times 
to suppress the impact and collect the average results. 
\qql{Programming language and network latency may impact the efficiency comparison among the tools: RIOLU was coded in Python, while FlashProfile was written in .NET, which is a more time-efficient language; ChatGPT inferred the patterns with powerful server hardware and ran remotely, while other tools were run locally. 
The time efficiency of RIOLU can be further improved with more efficient language or powerful hardware.} 

\noindent\underline{\textbf{Internal Validity. }}
We assume that the health data patterns should cover more records than anomaly patterns. Thus, 
we use K-Means to find a natural threshold for the patterns' coverage rate. 
\qql{K-Means is used given its low time complexity and wide use for data clustering. We acknowledge that more advanced clustering approaches (e.g., deep learning-based ones~\cite{ren2024deep}) exist; 
their 
combination with RIOLU can be verified in the future.} We recognize that the generated patterns' quality may be limited by the ignorance of data types (e.g., generating ``non-existing'' patterns for free text
)\qql{, or some records with rare legitimate patterns may be flagged due to their statistical minority}; we suggest that \st{this}\qql{these} problem\qql{s} could be feasibly fixed through human validation on the corresponding domain. 

\noindent\underline{\textbf{External Validity. }}
We tested RIOLU’s performance on data in various domains and sizes (e.g., the anomaly detection datasets have 999 to 100k records). 
However, our findings may be limited to the datasets that we considered.
Our work would benefit from future work that evaluates RIOLU on more diverse datasets, especially in practical settings.
\section{Conclusions}
\label{sec:conclusion_future_work}

This work proposes a fully automated pattern inference and anomaly detection approach, RIOLU, which does not require human labeling or parameter configuration. 
RIOLU leverages statistical heuristics and automated clustering to mitigate the problem of parameter configuration, and uses a rule-based progressive structure to ensure the precision of pattern inference. 
Our experiment results show that RIOLU can precisely generate patterns to describe data from various domains and detect pattern anomalies, in a fully automated manner, \heng{outperforming the baselines (including ChatGPT) in terms of both accuracy and efficiency}. 
Moreover, we found that Guided-RIOLU, a variant of RIOLU with the guidance of a small labeled dataset,  can further improve the anomaly detection performance. 
Our evaluation in an industrial setting proves the usefulness of our approach in a practical setting. RIOLU has the potential to benefit software and data engineering communities by improving the quality of the ever-growing data and the software built upon them. For example, our lightweight approach may be incorporated into the continuous integration pipeline of data-centric systems to support unit tests for data. 

\balance 

\bibliographystyle{IEEEtran}
\bibliography{references}

\end{document}